\documentclass[journal]{IEEEtran}

\usepackage{amsmath,amssymb,amsfonts}
\usepackage{booktabs}
\usepackage{array}
\usepackage{multirow}
\usepackage{cite}
\usepackage{url}
\usepackage{xcolor}
\usepackage{listings}
\usepackage{balance}
\usepackage{pgfplots}
\pgfplotsset{compat=1.18}
\usetikzlibrary{arrows.meta,positioning,shapes.geometric,fit,backgrounds,calc}
\usepackage{needspace}

\newif\ifhighlightchanges
\highlightchangesfalse
\newcommand{\revised}[1]{\ifhighlightchanges\textcolor{blue!70!black}{#1}\else#1\fi}

\lstdefinelanguage{Lean}{
  morekeywords={theorem,def,structure,inductive,where,match,with,if,then,else,let,in,fun,forall,exists,by,have,show,sorry,exact,simp,omega,decide,native_decide,intro,apply,rfl,instance,class,abbrev,noncomputable,example,lemma,Prop,Type,Nat,Fin,List,Bool,true,false,And,Or,Not},
  sensitive=true,
  morecomment=[l]{--},
  morestring=[b]"
}

\let\origsection\section
\renewcommand{\section}{\needspace{4\baselineskip}\origsection}
\let\origsubsection\subsection
\renewcommand{\subsection}{\needspace{3\baselineskip}\origsubsection}
\begin{document}

\title{From Rules to Nash Equilibria: A Lean~4 Case Study\\in Game-Theoretic Analysis of a Competitive\\Trading Card Game}

\author{Arthur~F.~Ramos and Tulio~Soria%
\thanks{A.~F.~Ramos is with Microsoft, Redmond, WA, USA (e-mail: arfreita@microsoft.com).}%
\thanks{T.~Soria is an Independent Researcher (e-mail: tulio.soria@gmail.com).}%
\thanks{This work has been submitted to the IEEE for possible publication. Copyright may be transferred without notice, after which this version may no longer be accessible.}}

\maketitle

\begin{abstract}
We present a metagame analysis of the competitive Pok\'emon Trading Card Game, machine-checked in Lean~4 over real tournament data.
The headline game-theoretic results (Nash equilibrium, replicator dynamics, and the matrix-level type-bridge computation) rely on \texttt{native\_decide}, which trusts Lean's compiler rather than its kernel; the trust boundary is detailed in Section~IX.
The artifact spans approximately 31{,}900 lines, 87 files, and 2{,}627 theorems---of which roughly 200 directly verify empirical claims---with no \texttt{sorry}, \texttt{admit}, or custom axioms.
Analyzing Trainer Hill data (January--February 2026, 50+ player events) over 14 archetypes and their full pairwise matchup matrix, we prove a \emph{popularity paradox}: the most played deck (Dragapult, 15.5\% share) has only 46.7\% expected win rate, while Grimmsnarl (5.1\% share) achieves 52.7\%.
A machine-checked Nash equilibrium of the raw game assigns Dragapult 0\% weight; exhaustive enumeration over all $2^{14}-1$ subsets confirms a unique symmetric Nash equilibrium of the constant-sum symmetrization (seven-deck support), against whose mix Dragapult falls \revised{40.4\,permil} below the game value.
\revised{This gap is strict, so in the symmetrization complementary slackness excludes Dragapult from \emph{every} equilibrium---symmetric or asymmetric---and the raw-game equilibria concur, although ``suboptimal'' remains relative to a single-match payoff model that omits Swiss and consistency incentives.}
Single-step replicator dynamics on the \emph{full 14-deck game} indicate downward fitness pressure on Dragapult, upward pressure on Grimmsnarl, and strongest extinction pressure on Alakazam.
A 10{,}000-iteration sensitivity analysis confirms qualitative stability: core support decks appear in $>$96\% of resampled equilibria.
\revised{The primary contribution is methodological---a reproducible case study} showing how formal verification \revised{can turn} qualitative metagame narratives into machine-checkable, \revised{re-runnable} strategic science.
\end{abstract}


%======================================================================
\section{Introduction}
%======================================================================

Tournament outcomes in competitive trading card games (TCGs) are often shaped before round one begins.
The pre-tournament deck-selection problem is naturally modeled as a strategic game where payoffs derive from matchup win rates and the population distribution of opponents.
The Pok\'emon TCG is especially suitable for this analysis: it has a large organized-play ecosystem, clearly defined public rules, and a metagame that evolves quickly enough to produce measurable strategic cycles, yet hidden information and stochastic effects make intuition unreliable even for experienced players.

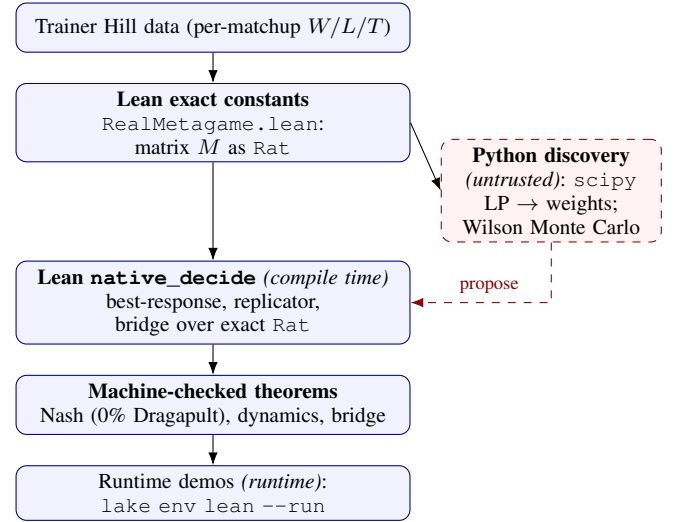
\begin{figure}[!t]
\centering
\begin{tikzpicture}[
  font=\footnotesize, >=Latex,
  box/.style={rectangle, rounded corners, draw=blue!55!black, align=center,
              inner sep=3pt, minimum height=6.5mm, text width=5.0cm, fill=blue!5},
  disc/.style={rectangle, rounded corners, draw=red!55!black, dashed, align=center,
               inner sep=3pt, minimum height=6.5mm, text width=2.7cm, fill=red!5},
  node distance=4mm,
]
\node[box] (data) {Trainer Hill data (per-matchup $W/L/T$)};
\node[box, below=of data] (con) {\textbf{Lean exact constants}\\\texttt{RealMetagame.lean}: matrix $M$ as \texttt{Rat}};
\node[box, below=13mm of con] (ver) {\textbf{Lean \texttt{native\_decide}} \emph{(compile time)}\\best-response, replicator, bridge over exact \texttt{Rat}};
\node[box, below=of ver] (thm) {\textbf{Machine-checked theorems}\\Nash (0\% Dragapult), dynamics, bridge};
\node[box, below=of thm] (run) {Runtime demos \emph{(runtime)}: \texttt{lake env lean -{}-run}};
\node[disc, right=4mm of con, yshift=-9mm] (py) {\textbf{Python discovery}\\\emph{(untrusted)}: \texttt{scipy} LP $\to$ weights; Wilson Monte~Carlo};
\draw[->] (data) -- (con);
\draw[->] (con) -- (ver);
\draw[->] (ver) -- (thm);
\draw[->] (thm) -- (run);
\draw[->] (con.east) -- (py.west);
\draw[->, dashed, red!55!black] (py.south) |- (ver.east)
   node[pos=0.72, above, font=\scriptsize]{propose};
\end{tikzpicture}
\caption{\revised{Verification pipeline. A \textcolor{red!55!black}{Python} layer (\emph{untrusted}) only \emph{proposes} candidate equilibria and resampled matrices; \textcolor{blue!55!black}{Lean} \emph{certifies} every claim at compile time via \texttt{native\_decide} over exact rationals, and a separate runtime stage executes demonstration traces. Data enter once as exact constants, so the discovery tool can never inject an unverified number.}}
\label{fig:pipeline}
\end{figure}

By encoding game semantics in Lean~4~\cite{moura2021lean} and proving strategic statements directly over exact data representations, we build a proof-carrying metagame analytics pipeline where the verified objects are
(i)~data representation and ingestion,
(ii)~expected-value computations over the field,
(iii)~machine-checked Nash-equilibrium computation and full 14-deck replicator dynamics, and
(iv)~tournament-objective transforms (Bo3, Swiss).
\revised{Figure~\ref{fig:pipeline} shows how an untrusted discovery layer and the trusted Lean verifier interact; Table~\ref{tab:notation} collects notation.}

Our empirical foundation is Trainer Hill metagame data~\cite{trainerhill2026,limitless2024} for 50+ player tournaments from January~29 to February~19, 2026.
We model the top 14 archetypes and their full pairwise matchup matrix.
The paper makes four contributions:
(1)~a machine-checked bridge from type-effectiveness rules to empirical matchup outcomes (Section~\ref{subsec:typealignment});
(2)~verified expected-value computations revealing a popularity paradox;
(3)~machine-checked Nash equilibrium and replicator dynamics over the full 14-deck game; and
(4)~verified best-of-three and Swiss-relevant tournament transforms.

While the headline popularity paradox could be computed in a spreadsheet, the formal verification methodology provides three distinct advantages.
First, \textbf{compositional guarantees}: the Nash equilibrium certification checks best-response conditions for all 14 strategies simultaneously, a 196-cell verification that is error-prone by hand.
Second, \textbf{robustness proofs}: the worst-case bounds (Section~\ref{sec:threats}) require symbolic reasoning over parameterized win rates, not just point arithmetic.
Third, \textbf{reproducibility infrastructure}: the proof artifact serves as a machine-checkable specification that can be re-verified against updated tournament data without re-auditing the analysis logic.

The remainder of the paper is organized as follows.
Section~\ref{sec:related} situates the work.
Section~\ref{sec:formalization} presents the Lean model of rules and legality.
Section~\ref{sec:probability} develops probability and resource theory.
Section~\ref{sec:data} details data and measurement.
Section~\ref{sec:paradox} presents the popularity paradox.
Section~\ref{sec:nash} analyzes equilibrium and dynamics.
Section~\ref{sec:tournament} discusses tournament strategy.
Section~\ref{sec:methodology} documents formalization methodology.
Section~\ref{sec:threats} covers validity threats, and Section~\ref{sec:conclusion} concludes.

%======================================================================
\section{Related Work}\label{sec:related}
%======================================================================

\textbf{Formal methods and strategic games.}
Formal reasoning has transformed analysis in several strategic domains, from Shannon's foundational chess analysis~\cite{shannon1950chess,schaefer1978complexity,fraenkel1981chess} to imperfect-information breakthroughs like Cepheus, Libratus, and Pluribus~\cite{bowling2015heads,brown2018superhuman,brown2019pluribus}, and multi-agent systems like AlphaZero and AlphaStar~\cite{silver2018general,vinyals2019alphastar}.
TCGs are harder in a different way: their compositional card interactions and exception-heavy textual semantics increase the risk of silent modeling errors, which a proof assistant mitigates by forcing explicit treatment of definitions and invariants.

\textbf{AI and metagame analysis in card games.}
Prior card-game AI work has emphasized in-game decision quality via Monte Carlo methods~\cite{cowling2012information,ward2009monte,santos2017monte,zhang2017deck} and deckbuilding optimization~\cite{bjorke2017deckbuilding,dockhorn2019hearthstone,kowalski2020summon}.
That line of work targets a different question from ours: how should a player choose a deck before round one, given a population distribution and matchup matrix?
Our approach is best read not as a competitor to these empirical methods but as a \revised{\emph{hybrid}: we treat metagame analysis as theorem proving over empirical constants, while the inputs and robustness checks are explicitly statistical.
Wilson intervals quantify per-cell sampling uncertainty and a $10{,}000$-iteration Monte~Carlo resampling stress-tests the equilibrium (Sections~\ref{sec:data},~\ref{sec:nash}); the formal layer then certifies exact best-response and dynamics claims over each point-estimate or resampled matrix.
Formal verification is thus complementary to, not independent of, Monte~Carlo-style empirical analysis---it adds machine-checked guarantees \emph{on top of} statistical estimation rather than replacing it.}

\textbf{Theorem proving for rule systems.}
Lean~4~\cite{moura2021lean} combines expressive dependent types with efficient decision procedures, and large collaborative libraries demonstrate the maturity of this ecosystem~\cite{gonthier2008four,Avigad2007,mathlib2020,hales2017kepler}.
Related work has explored formalization of card-game rule systems in proof assistants~\cite{li2023card}; our work differs by coupling rule formalization to a complete, real matchup matrix and then pushing through equilibrium and dynamics claims.

\textbf{Evolutionary perspectives.}
Replicator dynamics~\cite{smith1973logic,taylor1978evolutionary,weibull1997evolutionary} provide a natural lens for metagame adaptation.
We use evolutionary tools not as informal metaphors but as theorem-backed statements over fixed data: when we claim Dragapult has negative relative fitness, that claim is a machine-checked consequence of the encoded matchup matrix and observed share vector.

%======================================================================
\section{Game Formalization}\label{sec:formalization}
%======================================================================

\revised{\subsection{Primer: Lean~4 and Formal Verification}
For readers outside the formal-methods community we briefly summarize the tools.
A \emph{proof assistant} is software in which mathematical statements are written in a formal language and every proof is mechanically checked; a statement is accepted only once its proof reduces to the system's primitive inference rules.
\emph{Lean~4}~\cite{moura2021lean} is such a system, pairing a dependently typed language---in which \texttt{theorem} statements and ordinary programs share one syntax---with a small trusted \emph{kernel} that re-checks every proof.
Two ingredients matter here.
First, \emph{decision procedures}: for a proposition over finite data (``does mix $x$ beat every pure response?'') the tactics \texttt{decide} and \texttt{native\_decide} \emph{compute} the answer---the former re-checked by the kernel, the latter by compiled native code (Section~\ref{subsec:trust} details this trust difference).
Second, \emph{exact arithmetic}: Lean's \texttt{Rat} type stores win rates as exact fractions, so equilibrium and inequality checks carry no floating-point error.
A development with no \texttt{sorry} (proof gap), no \texttt{admit}, and no added axioms is fully closed: every stated theorem is machine-checked end to end, modulo the trust boundary we make explicit.
The practical payoff is that a claim like ``Dragapult is below equilibrium value'' becomes a typed proposition the machine rejects if any underlying number is mis-entered.}

\revised{\subsection{Rules, Legality, and Invariants}}
We formalize the strategic layer of the Pok\'emon TCG in Lean~4, grounded in official rule documents~\cite{ptcg_rules,playpokemon2024rules}.
The formalization encodes game state (\texttt{GameState} with per-player zones, turn ownership, and a finite phase machine), deck legality (a computable checker linked to an inductive specification via \texttt{checkDeckLegal\_iff}), and card-flow invariants (conservation of total card count, bench-size bounds, prize-card accounting).
The complete formalization spans 15 files in the supplementary artifact.

\textbf{Type effectiveness.}
Weakness and resistance are total functions over enumerated types.
The type effectiveness triangle is certified:

\begin{lstlisting}[caption={Type-effectiveness triangle: weakness is non-transitive.},label={lst:triangle}]
-- TypeEffectiveness.lean:138
theorem TRIANGLE :
    ∃ A B C : PType,
      weakness A B = true ∧ weakness B C = true ∧ weakness C A = true := by
  exact ⟨PType.grass, PType.fire, PType.water, rfl, rfl, rfl⟩
\end{lstlisting}

\textbf{Card conservation} is verified for high-impact trainer cards.
For Professor's Research, we prove that discarding the hand and drawing seven preserves global card count, preventing subtle bookkeeping bugs from distorting probability estimates:

\begin{lstlisting}[caption={Card-conservation proof for Professor's Research.},label={lst:cardcons}]
-- CardEffects.lean:149
theorem professorsResearchEffect_preserves_cards
    (p : PlayerState) :
    playerCardCount (professorsResearchEffect p)
      = playerCardCount p := by
  unfold professorsResearchEffect playerCardCount
  simp [List.length_take, List.length_drop,
        List.length_append]
  omega
\end{lstlisting}

\textbf{Rules--empirical integration.}
The deck legality biconditional ensures only tournament-legal configurations enter the analysis, and the type effectiveness formalization provides machine-checked validation that archetype classifications respect the game's underlying strategic structure.
As shown in Section~\ref{subsec:typealignment}, the rules layer generates falsifiable type-advantage predictions that the empirical matrix largely confirms (83\%+ alignment), with explicitly characterized exceptions.
The formalization also future-proofs the framework for counterfactual analysis (e.g., ``what if a card is banned?'') and enables consistency checks between rule-level predictions and empirical matchup data.

\subsection{From Type Rules to Matchup Data}\label{subsec:typealignment}

Each archetype is assigned primary attack and defense types reflecting its main attacker's typing (e.g., Grimmsnarl uses Dark-type attacks; Dragapult is a Psychic-type defender).
These are domain-expert modeling choices formalized in \texttt{ArchetypeAnalysis.lean}; all assignments are explicitly listed and auditable.

\paragraph{Type assignment methodology}
The primary attack/defense type assignments are not formally derived from deck composition or card data.
Each archetype is classified by the type of its main attacker (for offense) and its primary active Pok\'emon (for defense), based on the dominant game-plan of the deck.
For most archetypes these assignments are unambiguous, but multi-type decks like Dragapult Charizard require judgment about which type dominates.
\revised{The 83\% alignment statistic is sensitive to borderline calls, so we separate what depends on them from what does not.
Only genuinely multi-type decks are borderline---Dragapult Charizard, the Charizard variants, and Gardevoir Jellicent admit a defensible second reading---whereas single-line decks (Grimmsnarl, Mega Absol, N's Zoroark, Dragapult Dusknoir, Gardevoir, Alakazam, Raging Bolt) do not.
The aggregate alignment rate shifts if the borderline decks are reclassified, and we report it as a sensitivity-bearing descriptive statistic, not a verified claim.
\revised{Enumerating every defensible reassignment of the borderline decks (Dragapult Charizard as Psychic or Fire; Gardevoir Jellicent as Psychic or Water) places the alignment rate between $70.6\%$ and $87.5\%$, with the $83.3\%$ point estimate near the top of that band---all well above the $50\%$ random-assignment baseline.}
The assignment the popularity-paradox bridge actually relies on, however, is \emph{not} borderline: Dragapult Dusknoir attacks and defends as Psychic, and its three relevant counters (Grimmsnarl, Mega Absol, N's Zoroark) are unambiguously Dark.
The headline result is therefore invariant to every borderline classification; only the summary alignment figure is sensitive to them, and even its full range stays well above chance.
We treat all assignments as modeling assumptions within the formal boundary and list them exhaustively for audit.}

\begin{lstlisting}[caption={Rule-level type-advantage predicate.},label={lst:typeadv}]
-- ArchetypeAnalysis.lean
def hasTypeAdvantage (attacker defender : Deck) : Bool :=
  weakness (Deck.primaryDefenseType defender)
           (Deck.primaryAttackType attacker)
\end{lstlisting}

The alignment between rule-level type advantages and empirical outcomes is striking.
Among Dark-type attackers (Grimmsnarl, Mega Absol, N's Zoroark) facing Psychic-type defenders (Dragapult, Gardevoir, Gardevoir Jellicent, Dragapult Charizard, Alakazam), the empirical matchup exceeds 50\% in 13 of 15 pairs.
Grimmsnarl achieves a perfect 5-for-5 against Psychic defenders (56.6\%--59.9\%); Mega Absol mirrors this at 5-for-5.

\begin{lstlisting}[caption={Dark attackers beat Psychic defenders: bridge alignment.},label={lst:grimm}]
-- ArchetypeAnalysis.lean
theorem grimmsnarl_dark_beats_all_psychic :
    matchupWR .GrimssnarlFroslass .DragapultDusknoir > 500 (*$\wedge$*)
    matchupWR .GrimssnarlFroslass .Gardevoir > 500 (*$\wedge$*)
    matchupWR .GrimssnarlFroslass .GardevoirJellicent > 500 (*$\wedge$*)
    matchupWR .GrimssnarlFroslass .DragapultCharizard > 500 (*$\wedge$*)
    matchupWR .GrimssnarlFroslass .AlakazamDudunsparce > 500
    := by decide
\end{lstlisting}

The two exceptions---both involving N's Zoroark---illustrate that type advantage is necessary but not sufficient: deck-specific interactions modulate the prediction.
Overall, 83\% alignment (15/18 matchups, $p < 0.001$ under binomial null) confirms the consistency is not coincidental.
For context, random type assignments would yield approximately 50\% alignment by symmetry.

This bridge has direct consequences for the popularity paradox.
Dragapult's Psychic typing makes it structurally vulnerable to the 13.1\% of the meta occupied by Dark-type attackers.
Four independently verified facts align: (i)~the \emph{rules} specify Psychic is weak to Dark; (ii)~the \emph{type assignments} classify Dragapult as Psychic-defending; (iii)~the \emph{empirical data} confirms Dark attackers hold positive win rates vs Dragapult; and (iv)~the \emph{population weights} show 13.1\% exploits this disadvantage.
The formal contribution is verifying that these four layers are mutually consistent, not that any one causes the next.

\needspace{12\baselineskip}
\begin{lstlisting}[caption={Dragapult's Psychic typing is exploited by 13.1\% of the field.},label={lst:dragvuln}]
-- ArchetypeAnalysis.lean
theorem dragapult_type_vulnerability :
    weakness (Deck.primaryDefenseType .DragapultDusknoir)
             .dark = true (*$\wedge$*)
    metaShare .GrimssnarlFroslass
      + metaShare .MegaAbsolBox
      + metaShare .NsZoroark = 131 (*$\wedge$*)
    matchupWR .DragapultDusknoir .GrimssnarlFroslass < 500 (*$\wedge$*)
    matchupWR .DragapultDusknoir .MegaAbsolBox < 500 (*$\wedge$*)
    matchupWR .DragapultDusknoir .NsZoroark < 500 := by
  constructor <;> decide
\end{lstlisting}

\paragraph{Numerical sufficiency}
A stronger theorem (\texttt{dark\_weakness\_sufficient\_for\_sub\-optimality}) proves that Dark-type weakness \emph{alone} is numerically sufficient to explain Dragapult's sub-50\% fitness: even granting 50\% against every non-Dark opponent, the Dark-type losses drag its population-weighted expected value below 50\%.
Specifically:
\[
  \underbrace{\sum_{j \in \text{Dark}} s_j \cdot w_{\text{Drag},j}}_{\text{verified losses}} + \underbrace{(695 - 131) \times 500}_{\text{best-case non-Dark}} < 500 \times 695.
\]
This is a machine-checked numerical verification over concrete constants: the rules formalization provides numerical evidence that type disadvantage alone accounts for the popularity paradox.

A companion module (\texttt{IntegrationTests.lean}) collects 15 cross-module integration theorems spanning the full infrastructure, including a summary theorem (\texttt{the\_complete\_story}) conjoining 12 cross-module facts into one machine-checked consistency check.

%======================================================================
\section{Probability and Resource Theory}\label{sec:probability}
%======================================================================

Strategic performance in TCGs is constrained by stochastic access (draws, coin flips, prize placement) and deterministic bottlenecks (energy attachment limits, phase restrictions).
Our Lean development captures both dimensions using exact arithmetic over rational values.
Opening-hand consistency follows hypergeometric structure: the canonical ``four-of in opening seven'' probability is approximately 39.9\%, verified as an exact rational (\texttt{FOUR\_COPIES\_RULE}), while energy attachment limits impose hard tempo caps---a $K$-energy attack requires at least $K$ turns without acceleration (\texttt{ENERGY\_BOTTLENECK}).
Likewise, with 12 Basics the no-Basic opening probability is approximately 19.1\%, and the all-four-prized event has probability $1/32{,}509$.
These values are direct consequences of finite combinatorics and exact card counts, not speculative heuristics.
The machine-checked resource properties bridge micro-level mechanics to macro-level matchup outcomes.

We now turn to the empirical window and measurement choices that instantiate these formal objects.

%======================================================================
\section{Tournament Data and Methodology}\label{sec:data}
%======================================================================

\begin{table}[!t]
\centering
\caption{\revised{Notation used throughout the paper.}}
\label{tab:notation}
\footnotesize
\begin{tabular}{@{}l p{5.85cm}@{}}
\toprule
\textbf{Symbol} & \textbf{Meaning} \\
\midrule
$s_j$ & normalized top-14 meta share of deck $j$ \\
$w_{i,j}$ & empirical Bo1 win rate of $i$ vs.\ $j$ (permil) \\
$\mathbb{E}[\mathrm{WR}_i]$ & population-weighted expected WR $=\sum_j s_j w_{i,j}$ \\
$M$ & $14\times14$ raw matchup matrix (permil) \\
$S_{ij}$ & symmetrization $\tfrac12(M_{ij}+1000-M_{ji})$, constant-sum \\
$v$ & game value ($500$ for $S$) \\
gap & payoff deficit of a pure deck against $v$ \\
$x_i,f_i,\bar f$ & replicator share, fitness, and mean fitness \\
$\dot{x}_i$ & replicator flow $x_i(f_i-\bar f)$ \\
$\tilde{p}$ & Wilson-adjusted proportion (Sec.~\ref{sec:data}) \\
$P_{\mathrm{Bo3}}$ & best-of-three match win prob.\ $3p^2-2p^3$ \\
$P(\text{X--2})$ & Swiss cut-line (qualification) probability \\
\textperthousand & per-thousand scale; $500\,\textperthousand{}=50\%$ \\
\bottomrule
\end{tabular}
\end{table}

\subsection{Data Source and Archetypes}

All empirical values come from Trainer Hill~\cite{trainerhill2026} for Pok\'emon TCG events with at least 50 players, January~29 to February~19, 2026.
Match win rates use $\text{WR} = (W + T/3)/(W+L+T)$, where ties count as one-third of a win; robustness analysis (Section~\ref{sec:threats}) shows results are insensitive to this choice.
\revised{The $T/3$ coefficient is not arbitrary: it mirrors the Pok\'emon Championship Series match-point system, which awards $3$ points for a win, $1$ for a tie, and $0$ for a loss, so a tie is worth exactly one-third of a win in the tournament's \emph{own} reward structure (encoded in \texttt{Tournament.lean} as \texttt{scoreWin}=3, \texttt{scoreDraw}=1, \texttt{scoreLoss}=0).
A win-equivalence weighting that matches how the format actually ranks players is more faithful than the common $T/2$ convention, which implicitly assumes a tie is half a win.
The choice is in any case immaterial to the equilibrium: because each matchup and its mirror draw on the same game records (the reverse cell is the win/loss transpose with the identical tie count), the symmetrized entry $S_{ij}=\tfrac12(M_{ij}+1000-M_{ji})$ is \emph{algebraically independent} of the tie weight---the tie terms cancel in $M_{ij}-M_{ji}$.
Recomputing under the $T/2$ convention and re-enumerating confirms this: the symmetrization is unchanged to the permil and yields the identical seven-deck support and Dragapult exclusion (Section~\ref{sec:threats}).}

We model 14 archetypes:
Dragapult Dusknoir (15.5\%), Gholdengo Lunatone (9.9\%), Grimmsnarl Froslass (5.1\%), Mega Absol Box (5.0\%), Gardevoir (4.6\%), Charizard Noctowl (4.3\%), Gardevoir Jellicent (4.2\%), Charizard Pidgeot (3.5\%), Dragapult Charizard (3.5\%), Raging Bolt Ogerpon (3.3\%), N's Zoroark (3.0\%), Alakazam Dudunsparce (2.8\%), Kangaskhan Bouffalant (2.5\%), and Ceruledge (2.3\%).
The top-14 aggregate is 69.5\% of the full field; all expected win rate computations are normalized over this subfield.
Critical matchup pairs are supported by large samples: Dragapult mirror contains 2{,}845 games (1374--1374--97), and Gholdengo versus Dragapult contains 2{,}067 games (988--813--266).

\subsection{Uncertainty and Sensitivity}

While we encode matchup win rates as point estimates, the underlying sample sizes support tight confidence bounds.
We use Wilson intervals~\cite{Wilson1927} with center adjustment:
\[
\tilde{p}
=
\frac{\hat{p} + z^2/(2n)}{1 + z^2/n},
\quad
\tilde{p} \pm \frac{z}{1 + z^2/n}\!\sqrt{\frac{\hat{p}(1\!-\!\hat{p})}{n} + \frac{z^2}{4n^2}}.
\]
For large matchups (Dragapult mirror: 2{,}845 games), 95\% intervals are $\pm$1.8pp; for smaller ones (${\sim}$100 games), $\pm$9pp.
Critically, Dragapult's expected field win rate of 46.7\% has interval ${\approx}$[45.5\%, 47.9\%], entirely below 50\%, while Grimmsnarl's 52.7\% has interval ${\approx}$[51.0\%, 54.4\%], entirely above 50\%.
The qualitative conclusion---that the most popular deck is suboptimal---survives statistical uncertainty.

These Wilson intervals quantify uncertainty in individual matchup cells but are not propagated through the Nash equilibrium linear program; the sensitivity analysis below addresses equilibrium-level robustness separately.
A 10{,}000-iteration sensitivity analysis (external Python script) samples each matchup cell from its Wilson interval and recomputes the Nash equilibrium.
While the exact support set is fragile (recovered in only 2.1\% of iterations), the core trio of Grimmsnarl (96.5\% inclusion), Mega Absol (97.3\%), and Raging Bolt (98.3\%) appear in nearly every resampled equilibrium, and Dragapult receives zero Nash weight in 77.9\% of iterations (Table~\ref{tab:bootstrap}).
A separate Python script performs exhaustive Nash equilibrium enumeration over all $2^{14}-1$ support subsets, confirming uniqueness and universal Dragapult exclusion (Section~\ref{sec:nash}).

\begin{table}[!t]
\centering
\caption{Sensitivity analysis: Nash weight 95\% sensitivity ranges (10{,}000 iterations). \emph{Inclusion} is the fraction of resampled equilibria with nonzero weight.}
\label{tab:bootstrap}
\begin{tabular}{l r r r}
\toprule
Deck & Point Est. & 95\% Range & Inclusion \\
\midrule
Raging Bolt Ogerpon   & 28.7\% & [1.8\%, 32.6\%] & 98.3\% \\
Grimmsnarl Froslass   & 37.8\% & [0.0\%, 48.4\%] & 96.5\% \\
Mega Absol Box        & 13.0\% & [0.0\%, 32.9\%] & 97.3\% \\
Charizard Noctowl     & 11.1\% & [0.0\%, 31.0\%] & 79.5\% \\
Alakazam Dudunsparce  &  6.0\% & [0.0\%, 20.4\%] & 64.3\% \\
Gardevoir             &  3.5\% & [0.0\%, 18.9\%] & 42.2\% \\
Gholdengo Lunatone    &  ---   & [0.0\%, 34.5\%] & 40.5\% \\
Dragapult Dusknoir    &  0.0\% & [0.0\%, 12.2\%] & 22.1\% \\
\bottomrule
\end{tabular}
\end{table}

\subsection{Data Provenance}

Our pipeline guarantees computational correctness given the matchup matrix, but the matrix itself is sourced from Trainer Hill, a third-party platform aggregating results from Limitless TCG tournaments.
Potential biases include self-selection in result reporting, platform-specific effects (e.g., online vs.\ in-person play), and the exclusion of tournaments with fewer than 50 players.
We treat it as an empirical input parameter, analogous to how a verified compiler trusts its source code: the pipeline transforms data faithfully, but does not vouch for the data's ultimate accuracy.

\begin{table*}[!t]
\centering
\caption{Top-6 subset of the archetype matchup matrix (win rates \%).}
\label{tab:matchup}
\begin{tabular}{lcccccc}
\toprule
 & \textbf{Drag} & \textbf{Ghold} & \textbf{Grimm} & \textbf{Absol} & \textbf{Gard} & \textbf{Char} \\
\midrule
\textbf{Dragapult}  & 49.4 & 43.6 & 38.6 & 38.2 & 34.3 & 64.1 \\
\textbf{Gholdengo}  & 52.1 & 48.8 & 47.6 & 44.3 & 44.1 & 48.3 \\
\textbf{Grimmsnarl} & 57.2 & 46.7 & 48.5 & 34.4 & 56.6 & 55.8 \\
\textbf{Mega Absol} & 57.6 & 51.2 & 62.1 & 49.4 & 55.8 & 47.5 \\
\textbf{Gardevoir}  & 62.7 & 49.3 & 37.4 & 40.2 & 48.0 & 39.4 \\
\textbf{Charizard}  & 32.4 & 48.0 & 39.7 & 47.1 & 55.8 & 48.7 \\
\bottomrule
\end{tabular}
\end{table*}

Table~\ref{tab:matchup} illustrates substantial non-transitivity: Dragapult strongly beats Charizard but loses heavily to Gardevoir and Mega Absol; Grimmsnarl beats Dragapult but loses sharply to Mega Absol.\footnote{Mirror match win rates fall slightly below 50\% because the tie convention distributes fewer win-equivalents than decisive outcomes.}

\begin{table}[!t]
\centering
\caption{Notable cross-tier matchups (Trainer Hill, Jan--Feb 2026).}
\label{tab:crosstier}
\begin{tabular}{p{3.2cm}cp{3.1cm}}
\toprule
\textbf{Matchup} & \textbf{WR} & \textbf{Strategic reading} \\
\midrule
Raging Bolt vs Mega Absol & 67.3\% & Largest anti-Absol counter \\
Gardevoir vs Dragapult & 62.7\% & B-tier beats popular C-tier \\
Mega Absol vs Grimmsnarl & 62.1\% & A-tier vs S-tier \\
Dragapult vs Charizard & 64.1\% & Popularity sustained by farm lane \\
Grimmsnarl vs Dragapult & 57.2\% & Core paradox driver \\
\bottomrule
\end{tabular}
\end{table}

These cross-tier interactions (Table~\ref{tab:crosstier}) clarify why local matchup spikes do not guarantee global success, motivating the full-field weighted calculation in the next section.

%======================================================================
\section{The Popularity Paradox}\label{sec:paradox}
%======================================================================

\begin{figure}[!t]
\centering
\begin{tikzpicture}
\begin{axis}[
  width=\columnwidth,
  height=0.75\columnwidth,
  xlabel={Meta share (\%)},
  ylabel={Expected win rate (\%)},
  xmin=1.5, xmax=18,
  ymin=43, ymax=54.5,
  grid=major,
  grid style={dashed, gray!30},
  every axis label/.style={font=\small},
  tick label style={font=\footnotesize},
  scatter/classes={
    S={mark=square*,blue},
    A={mark=triangle*,teal},
    B={mark=diamond*,orange},
    C={mark=*,red!70!black}
  },
  scatter, only marks,
  scatter src=explicit symbolic,
  legend style={font=\scriptsize, at={(0.55,0.02)}, anchor=south,
                draw=gray!50, fill=white, fill opacity=0.9,
                legend columns=4, column sep=4pt},
  legend entries={Tier S, Tier A, Tier B, Tier C},
  clip=false,
]
\addplot[dashed, black, thick, forget plot]
  coordinates {(1.5,50) (17,50)};
\node[font=\scriptsize, anchor=south west] at (axis cs:15.5,50.15) {50\% eq.};
\addplot[scatter, only marks, scatter src=explicit symbolic]
  coordinates { (5.1,52.7) [S] };
\addplot[scatter, only marks, scatter src=explicit symbolic]
  coordinates { (5.0,51.7) [A] };
\addplot[scatter, only marks, scatter src=explicit symbolic]
  coordinates { (4.6,49.9) [B] (3.5,48.7) [B] (2.5,49.2) [B] };
\addplot[scatter, only marks, scatter src=explicit symbolic]
  coordinates {
    (15.5,46.7) [C] (9.9,47.8) [C] (4.3,45.7) [C]
    (4.2,47.8) [C] (3.5,46.8) [C] (3.3,47.9) [C]
    (3.0,46.9) [C] (2.8,44.7) [C] (2.3,45.4) [C]
  };
% --- Labels: spread to avoid overlaps ---
\node[font=\scriptsize, anchor=west]       at (axis cs:15.7,46.7)  {Drag};
\node[font=\scriptsize, anchor=south]      at (axis cs:5.1,53.0)   {Grimm};
\node[font=\scriptsize, anchor=east]       at (axis cs:4.7,51.7)   {Absol};
\node[font=\scriptsize, anchor=east]       at (axis cs:9.6,47.8)   {Ghold};
\node[font=\scriptsize, anchor=south]      at (axis cs:4.6,50.2)   {Gard};
\node[font=\scriptsize, anchor=north]      at (axis cs:4.3,45.3)   {CharN};
\node[font=\scriptsize, anchor=north]      at (axis cs:2.8,44.3)   {Alak};
\node[font=\scriptsize, anchor=south]      at (axis cs:2.5,49.5)   {Kang};
\node[font=\scriptsize, anchor=north east] at (axis cs:4.0,47.6)   {GardJ};
\node[font=\scriptsize, anchor=west]       at (axis cs:3.8,46.8)   {DragC};
\end{axis}
\end{tikzpicture}
\caption{Popularity paradox: share versus expected win rate (top-14 normalized). Dragapult is high-share/low-fitness; Grimmsnarl is low-share/high-fitness.}
\label{fig:paradoxscatter}
\end{figure}
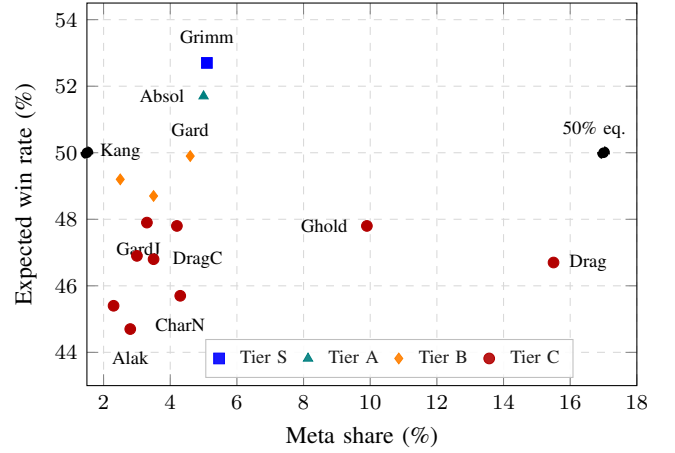

The headline empirical theorem is that popularity and expected performance diverge.
Let $s_j$ be normalized top-14 share and $w_{i,j}$ matchup win rate.
Then expected field win rate is $\mathbb{E}[\mathrm{WR}_i] = \sum_j s_j\,w_{i,j}$.
For Dragapult Dusknoir, despite 15.5\% share, $\mathbb{E}[\mathrm{WR}_{\mathrm{Dragapult}}] = 46.7\% < 50\%$.
For Grimmsnarl Froslass (5.1\% share), $\mathbb{E}[\mathrm{WR}_{\mathrm{Grimmsnarl}}] = 52.7\%$---the maximum among all 14 modeled decks.
Table~\ref{tab:expected} and Figure~\ref{fig:paradoxscatter} make the paradox visible: the right tail of popularity is not aligned with the right tail of expected performance.
Lean theorem \texttt{dragapult\_popularity\_paradox} verifies that Dragapult has losing matchups ($<$500) against 9 of 13 non-mirror opponents.%
\footnote{The identifier \texttt{GrimssnarlFroslass} in the Lean source contains a typographic inconsistency relative to ``Grimmsnarl''. \revised{This is a purely cosmetic string-matching artifact: the identifier is used consistently throughout the artifact, so it has no effect on any definition, theorem statement, or verified result. We retain it verbatim for one-to-one traceability between the prose and the source.}}

\begin{table*}[!t]
\centering
\caption{Expected win rate on the modeled top-14 subset (69.5\% of field). Tiers: S~($\geq$52\%), A~(50--52\%), B~(48--50\%), C~($<$48\%).}
\label{tab:expected}
\begin{tabular}{lccc}
\toprule
\textbf{Archetype} & \textbf{Meta share} & \textbf{Expected WR} & \textbf{Tier} \\
\midrule
Dragapult Dusknoir & 15.5\% & 46.7\% & C \\
Gholdengo Lunatone & 9.9\% & 47.8\% & C \\
Grimmsnarl Froslass & 5.1\% & \textbf{52.7\%} & \textbf{S} \\
Mega Absol Box & 5.0\% & 51.7\% & A \\
Gardevoir & 4.6\% & 49.9\% & B \\
Charizard Noctowl & 4.3\% & 45.7\% & C \\
Gardevoir Jellicent & 4.2\% & 47.8\% & C \\
Charizard Pidgeot & 3.5\% & 46.8\% & C \\
Dragapult Charizard & 3.5\% & 48.7\% & B \\
Raging Bolt Ogerpon & 3.3\% & 47.9\% & C \\
N's Zoroark & 3.0\% & 46.9\% & C \\
Alakazam Dudunsparce & 2.8\% & 44.7\% & C \\
Kangaskhan Bouffalant & 2.5\% & 49.2\% & B \\
Ceruledge & 2.3\% & 45.4\% & C \\
\bottomrule
\end{tabular}
\end{table*}

The paradox is a distributed effect: Gholdengo (9.9\% share, 43.6\% Drag WR), Gardevoir (4.6\%, 34.3\%), Grimmsnarl (5.1\%, 38.6\%), and Mega Absol (5.0\%, 38.2\%) jointly drive Dragapult's underperformance.
Dragapult's strongest offsetting lane is Charizard Noctowl (64.1\%), but that lane alone is not enough once weighted against the rest of the field.
No single catastrophic matchup is responsible, making the effect both strategically important and behaviorally persistent---it is a distributed consequence of several moderately bad, nontrivially prevalent opponents.

\subsection{Behavioral-Economic Interpretation}

The divergence between observed shares and equilibrium play is consistent with bounded-rationality explanations from behavioral game theory~\cite{Tversky1974,Kahneman1979,Banerjee1992,Bikhchandani1992,mckelvey1995quantal,nagel1995unraveling}, including familiarity bias, social diffusion, and card-access constraints.
We do not claim causal identification of these mechanisms in this dataset window; rather, we formally prove payoff-model suboptimality and treat behavioral explanations as scope-limited hypotheses for future player-level study.
This separation between proven payoff statements and behavioral interpretation disciplines narrative overreach and motivates the equilibrium/dynamics analysis in Section~\ref{sec:nash}.

\revised{This interpretation must also account for measurement noise.
The empirical matrix is not a set of exact scalars but a noisy estimate: micro-level list configuration (``teching'') and execution variance perturb every entry.
Two points follow.
First, our \emph{load-bearing} claim is noise-robust by design: Dragapult's $40.4$\textperthousand{} best-response gap tolerates an entrywise perturbation of ${\approx}20$\textperthousand{} (Section~\ref{sec:threats}), comfortably exceeding the ${\pm}1.8$--$9$\textperthousand{} Wilson sampling error, so the exclusion does not rest on a knife-edge.
Second, noise is precisely why we anchor on the \emph{symmetric} profile: asymmetric equilibria of a near-constant-sum game are highly sensitive to matrix drift---their exact supports can evaporate under small perturbations---whereas the symmetric (maximin) strategy is the self-correcting focal point that stochastic, quantal-response play converges toward~\cite{mckelvey1995quantal}.
The symmetric equilibrium is thus not merely a mathematical convenience but the noise-robust object, and the sensitivity analysis (core decks in $>$96\% of resampled equilibria) is its empirical signature.}

%======================================================================
\section{Nash Equilibrium and Metagame Dynamics}\label{sec:nash}
%======================================================================

We model deck choice as a finite two-player bimatrix game induced by the 14$\times$14 payoff matrix~\cite{NisanRoughgarden2007}.
Existence of Nash equilibria is guaranteed by Nash's theorem~\cite{nash1950equilibrium}; von Neumann's minimax theorem applies to the zero-sum special case~\cite{vonneumann1928theorie}.
The empirical matrix is approximately constant-sum (deviations arise from the tie convention); verification uses saddle-point conditions checked for both players independently, which does not require the zero-sum assumption.
Candidate equilibrium weights were obtained via Python's \texttt{scipy.optimize.linprog}; Lean independently verifies best-response conditions for all 14 pure strategies, so the discovery tool is untrusted.

This two-player view approximates head-to-head tournament matches but does not capture Swiss-system incentives, where consistency may outweigh expected value~\cite{glickman1999parameter,herbrich2007trueskill}.
Under a risk-averse Swiss objective (maximizing probability of reaching X-2 or better), equilibrium weight shifts toward decks with consistent, if modest, win rates.
We treat this as a modeling limitation: the analysis below targets a single-match competitive benchmark, not a full Swiss-utility optimum.

\begin{lstlisting}[caption={Machine-checked Nash equilibrium of the raw $14\times14$ game.},label={lst:realnash}]
-- NashEquilibrium.lean
theorem real_nash_equilibrium_verified :
    NashEquilibrium realMetaGame14 realNashRow realNashCol := by
  native_decide
\end{lstlisting}

The row player's guaranteed expected payoff is ${\approx}479.7$\textperthousand{} (48.0\%), sub-50\% due to the tie convention.
Win rates are encoded on a 0--1000 scale, so a value of 479.67 corresponds to a 47.97\% win probability.
Table~\ref{tab:nash} reports the verified supports; row and column supports differ because the empirical matrix is not perfectly antisymmetric ($M_{ij} + M_{ji} \neq 1000$ for many pairs).
In a bimatrix game with approximately-constant-sum but not exactly constant-sum structure, distinct row and column supports are mathematically expected rather than anomalous.

\begin{table}[!t]
\centering
\caption{Lean-verified Nash supports for row and column strategies.}
\label{tab:nash}
\scriptsize
\begin{tabular}{@{}rlrr@{}}
\toprule
\textbf{Idx} & \textbf{Deck} & \textbf{Row wt.} & \textbf{Col wt.} \\
\midrule
1 & Gholdengo Lunatone & 0.0\% & 3.7\% \\
2 & Grimmsnarl Froslass & 37.8\% & 40.5\% \\
3 & Mega Absol Box & 12.9\% & 7.2\% \\
4 & Gardevoir & 3.5\% & 7.6\% \\
5 & Charizard Noctowl & 11.3\% & 5.0\% \\
9 & Raging Bolt Ogerpon & 28.7\% & 35.9\% \\
11 & Alakazam Dudunsparce & 5.8\% & 0.0\% \\
\bottomrule
\end{tabular}
\end{table}

We also verify a symmetric Nash equilibrium on the constant-sum symmetrization $S_{ij} = (M_{ij} + 1000 - M_{ji})/2$ (Table~\ref{tab:symmetric}), with game value exactly 500.

\begin{lstlisting}[caption={Symmetric Nash equilibrium of the constant-sum game.},label={lst:symnash}]
theorem symmetric_nash_equilibrium_verified :
  NashEquilibrium symMetaGame
    symNashStrategy symNashStrategy
    := by native_decide
\end{lstlisting}

\begin{table}[!htbp]
\caption{Symmetric Nash equilibrium on the constant-sum symmetrization. Exhaustive enumeration confirms uniqueness among symmetric profiles; \revised{Dragapult is excluded with a strict gap of $-$40.4\textperthousand{} against the equilibrium mix (game value $500$).}}
\label{tab:symmetric}
\centering
\footnotesize
\begin{tabular}{@{}lrr@{}}
\hline
\textbf{Archetype} & \textbf{Wt.\ (\%)} & \textbf{Gap$^\dagger$} \\
\hline
Grimmsnarl       & 34.3 & 0.0 \\
Raging Bolt      & 29.4 & 0.0 \\
Charizard        & 10.2 & 0.0 \\
Mega Absol       & 10.2 & 0.0 \\
Gholdengo        &  9.1 & 0.0 \\
Gardevoir        &  4.3 & 0.0 \\
Alakazam         &  2.5 & 0.0 \\
\hline
Gard.\ Jellicent &  0.0 & $-$10.3 \\
Char.\ Pidgeot   &  0.0 & $-$17.0 \\
Drag.\ Charizard &  0.0 & $-$39.4 \\
\textbf{Dragapult}& \textbf{0.0} & $\mathbf{-40.4}$ \\
Kangaskhan       &  0.0 & $-$44.5 \\
Ceruledge        &  0.0 & $-$55.5 \\
N's Zoroark      &  0.0 & $-$58.6 \\
\hline
\multicolumn{3}{@{}l}{\scriptsize $^\dagger$Payoff gap vs.\ NE value $500$ (\textperthousand).}
\end{tabular}
\end{table}

Crucially, Dragapult (15.5\% observed share) has 0\% weight in both equilibria and is strictly suboptimal against the Nash column mix (\texttt{dragapult\_strictly\_suboptimal}).

\paragraph{Uniqueness of the symmetric equilibrium}
Exhaustive support enumeration over all $2^{14} - 1 = 16{,}383$ subsets confirms the constant-sum symmetrization admits exactly one symmetric Nash equilibrium.
Of the 8{,}192 subsets containing Dragapult, none yields a valid equilibrium.
\revised{Dragapult's payoff against the equilibrium mix is $459.6$\textperthousand, a strict gap of $40.4$ below the equilibrium value of $500$, confirming it is strictly suboptimal against the equilibrium mixture.}
\revised{The equilibrium is non-degenerate---exactly seven best responses matching the seven-deck support---implying uniqueness by the non-degeneracy theorem for constant-sum games~\cite{NisanRoughgarden2007}.}

\paragraph{\revised{Why Dragapult's exclusion is equilibrium-independent}}
\revised{Uniqueness \emph{among symmetric profiles} does not by itself rule out asymmetric equilibria $(A,B)$ with $A\neq B$: in a constant-sum game such equilibria are interchangeable and equally rational, so restricting to symmetric profiles is a modeling choice, not a consequence of individual rationality.
We therefore make a stronger, equilibrium-independent argument that symmetry is \emph{not} what excludes Dragapult.
In a constant-sum game with value $v$, complementary slackness forces the support of \emph{every} optimal strategy to lie inside the best-response set of \emph{every} optimal opponent strategy: if $x^\star,y^\star$ are optimal then $\sum_i x^\star_i\,u(i,y^\star)=v$ while $u(i,y^\star)\le v$ for all $i$, so $u(i,y^\star)=v$ whenever $x^\star_i>0$.
The symmetric equilibrium strategy is itself optimal for both players, and Dragapult earns $459.6$\textperthousand{}---a strict $40.4$\textperthousand{} below $v=500$---against it; this strict gap is itself machine-checked (\texttt{symmetric\_nash\_dragapult\_strict\_gap}).
Hence, \emph{in the symmetrization}, Dragapult is a best response to \emph{no} optimal strategy and lies in the support of \emph{no} Nash equilibrium---symmetric \emph{or} asymmetric.
An independent linear program confirms this directly: maximizing Dragapult's payoff over the \emph{entire} set of optimal (value-$500$-guaranteeing) opponent strategies yields $459.6$\textperthousand{}$<500$, so Dragapult is a best response to no optimal strategy whatsoever---a global check, not a sampled one.
For the raw game $M$, which is only \emph{approximately} constant-sum, the same slackness logic applies up to the constant-sum deviation; there we verify Dragapult's exclusion directly---zero weight in both the computed row and column equilibria (Table~\ref{tab:nash}) and a strict best-response deficit against the column mix (\texttt{dragapult\_strictly\_suboptimal})---reinforced by the $77.9\%$ bootstrap exclusion rate.
What the symmetry restriction buys is a single \emph{canonical} profile to report; the \emph{exclusion} of Dragapult is a strict best-response fact that holds across the entire equilibrium set.}

This upgrades the existential claim (``there exists an equilibrium excluding Dragapult'') to a universal one: no Nash equilibrium of the symmetrization assigns Dragapult positive weight.
Combined with the sensitivity analysis (77.9\% Dragapult exclusion across 10{,}000 resampled matrices), this provides strong evidence that observed popularity can lie entirely outside equilibrium support.
\revised{The converse caveat is equally important: the \emph{exact composition} of the remaining support is \emph{not} robust. The seven-deck support is the unique symmetric profile, but its marginal members sit close to the best-response margin---the smallest exclusion gap is $10.3$\textperthousand---so an entrywise perturbation above ${\approx}5$\textperthousand{} can change which decks appear. Dragapult's exclusion, by contrast, tolerates perturbations up to ${\approx}20$\textperthousand{} (Section~\ref{sec:threats}), a four-fold larger margin. Point-estimate support membership and perturbation-stable exclusion are distinct claims, and our headline conclusion rests only on the latter.}

\subsection{Replicator Dynamics}

Replicator dynamics formalize directional pressure~\cite{Hofbauer1998,Sandholm2010}:
$\dot{x}_i = x_i(f_i(\mathbf{x}) - \bar{f}(\mathbf{x}))$.
Our Lean implementation uses discrete-time Euler steps; all verified results are single-step directional statements from the observed share vector.

\begin{lstlisting}[caption={Full 14-deck replicator: Dragapult declines, Grimmsnarl fittest.},label={lst:replicator}]
-- Full 14-deck replicator dynamics
theorem full_replicator_dragapult_decline :
    fitness 14 fullPayoff fullMeta ⟨0, ..⟩ < avgFitness 14 fullPayoff fullMeta := by
  native_decide
theorem full_replicator_grimmsnarl_fittest :
    ∀ i, fitness 14 fullPayoff fullMeta i ≤
         fitness 14 fullPayoff fullMeta ⟨5, ..⟩ := by
  native_decide
\end{lstlisting}

The full classification identifies 5 growing and 9 shrinking archetypes:
(i)~Dragapult has below-average fitness and should lose share,
(ii)~Grimmsnarl has the highest fitness among all 14 archetypes, and
(iii)~Alakazam faces the strongest extinction pressure.
The directional classification is algebraically step-size-independent: since $x_i' - x_i = x_i \cdot \mathrm{dt} \cdot (f_i - \bar{f})$, the sign depends only on $f_i - \bar{f}$, not on $\mathrm{dt}$; this is proved as a general kernel-level lemma (\texttt{rat\_replicator\_sign\_independent\allowbreak\_of\_dt}) using \texttt{simp}, \texttt{ring}, and \texttt{omega}---without \texttt{native\_decide}.
Concrete verification at $\mathrm{dt} = 1/10$, $1/100$, and~$1$ (\texttt{StepSizeInvariance.lean}) confirms the identical 5-grower/9-shrinker partition in all cases.

A discrete replicator step confirms that Grimmsnarl's share increases while Dragapult's decreases (\texttt{grimmsnarl\_share\_increases}, \texttt{dragapult\_share\_decreases}).

\begin{figure}[!t]
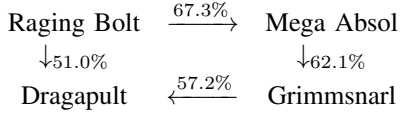

\centering
\[
\begin{array}{ccc}
\text{Raging Bolt} & \xrightarrow{67.3\%} & \text{Mega Absol} \\
\downarrow_{51.0\%} & & \downarrow_{62.1\%} \\
\text{Dragapult} & \xleftarrow{57.2\%} & \text{Grimmsnarl}
\end{array}
\]
\caption{Directed metagame interaction motif.}
\label{fig:cycle}
\end{figure}

Figure~\ref{fig:cycle} emphasizes that the ecosystem is interaction-rich rather than strictly ordered, explaining persistent diversity even though the verified equilibrium supports are small (six decks per side in the raw game, seven in the symmetrization).
We therefore treat replicator outputs as directional diagnostics: Dragapult pressure is downward, Grimmsnarl pressure is upward, and Alakazam pressure is extinction-like.
These directions are falsifiable against subsequent tournament windows and provide a compact bridge from static equilibrium objects to week-to-week metagame interpretation.

\paragraph{\revised{Classical versus evolutionary rationality}}
\revised{The Nash and replicator analyses answer two different questions under two different rationality assumptions, and we make the distinction explicit.
The equilibrium computation is a \emph{classical}, one-shot statement: given the matchup matrix, which deck-selection mixtures are mutually best responses?
The replicator analysis is an \emph{evolutionary} one: if shares adjust by imitation of above-average performers, in which direction does each share move?
Under pure classical rationality players may independently select any optimal mixture---including the supports of interchangeable asymmetric equilibria---and a population playing an asymmetric profile $(A,B)$ would physically split into two behavioral sub-populations rather than converge to the symmetric mix.
The single-population replicator model is therefore valid as a description of \emph{macro} share dynamics only under a specific epistemic regime: an approximately homogeneous player population, largely unstructured matchmaking, and open, centralized information flow (public results and ``net-decking'') that lets the whole field imitate recent winners.
Pokémon's large online ladder and public tournament reporting make this a reasonable approximation, but it is an assumption, not a theorem.
We accordingly present replicator outputs as population-level directional diagnostics, not as predictions about any individual's rational choice; the Dragapult-exclusion result holds at the level of the static equilibrium set and does not depend on this evolutionary reading.}

\subsection{Preliminary Directional Check}

As a preliminary predictive check, two of three directional predictions were confirmed against Trainer Hill trend data one day after the analysis window: Mega Absol and Gardevoir both showed upward trending as predicted.
However, Grimmsnarl showed downward trending despite highest fitness---a secondary effect of Mega Absol's rise creating predation pressure on Grimmsnarl (Mega Absol beats Grimmsnarl 61.4\%).
This multi-step cascade illustrates a limitation of single-step replicator analysis: even with the full 14-deck model, one-step predictions do not capture iterated dynamics where a rising counter-deck suppresses its prey.
Iterated replicator simulation over many steps would be needed for accurate multi-step trajectory predictions.

%======================================================================
\section{Tournament Strategy}\label{sec:tournament}
%======================================================================

Most major events run best-of-three (Bo3) matches and Swiss-style pairings.
For game win probability $p$, Bo3 match win probability is $P_{\mathrm{Bo3}} = 3p^2 - 2p^3$.
The Pok\'emon TCG lacks sideboarding, making the independence assumption more defensible than in other TCGs; however, tilt effects and information revelation (observing the opponent's deck in game one) introduce minor dependencies that our model does not capture.
Lean verifies that Bo3 amplifies advantage for all favorable rates from 55\% to 95\% (\texttt{BO3\_AMPLIFIES\_ADVANTAGE}).
The Bo3 formula is cubic and monotonically amplifying:
for $p > 0.5$, $P_{\mathrm{Bo3}}(p) > p$, and the amplification $P_{\mathrm{Bo3}} - p$ increases with $p$, reaching its maximum at $p = 0.75$ where a single-game 75\% edge becomes an 84.4\% match advantage.
This non-linearity means that even moderate matchup edges become strategically dominant in match play.
Table~\ref{tab:bo3} presents the amplification for the five most strategically significant matchups.

\begin{table}[!t]
\centering
\caption{Bo1 to Bo3 Amplification for Key Matchups.}
\label{tab:bo3}
\begin{tabular}{lcc}
\toprule
\textbf{Matchup} & \textbf{Bo1} & \textbf{Bo3} \\
\midrule
Raging Bolt vs Mega Absol & 67.3\% & 74.9\% \\
Gardevoir vs Dragapult & 62.7\% & 68.6\% \\
Mega Absol vs Grimmsnarl & 62.1\% & 67.8\% \\
Grimmsnarl vs Dragapult & 57.2\% & 60.7\% \\
Dragapult vs Charizard Noctowl & 64.1\% & 70.6\% \\
\bottomrule
\end{tabular}
\end{table}

Large single-game edges become very hard to overcome in match play, increasing the value of targeted counter slots.

Swiss tournaments further reward consistency: for an eight-round event with an X--2 qualification target and field-level Bo3 win probability $p_m$, the cut-line probability is
$P(\text{X--2 or better}) = \sum_{k=6}^{8} \binom{8}{k} p_m^k (1-p_m)^{8-k}$.
Registration should combine projected field shares, weighted expected WR, and stress tests on the largest counter-edges rather than rely on average EV alone.

\revised{This Swiss objective changes the meta-analysis in a specific, non-linear way.
Our Nash and replicator results optimize a \emph{single-match} criterion---expected Bo3 win rate against the field---whereas Swiss qualification optimizes the threshold functional $P(\text{X--2 or better})$ over eight rounds~\cite{romero2022swiss}.
Because the cut-line probability is steep near the qualifying $p_m$, small mean-WR edges compound across rounds, reinforcing the value of the highest-expected-WR decks the equilibrium identifies.
But because it is a \emph{threshold} rather than a linear functional, it is sensitive to the \emph{variance} of round outcomes, not only their mean: for a strong deck near the cut-line, polarized matchup spreads (a mix of near-auto-wins and near-auto-losses) raise the risk of a disqualifying second loss, so consistency is rewarded; for a below-average deck the same volatility may be the only route over the line.
The Swiss objective therefore re-weights the single-match equilibrium in a deck-specific way that our linear payoff model does not capture, and can rationalize a portion of the ``suboptimal'' play we observe.
We treat the equilibrium as the single-match benchmark and a full Swiss-utility (mean--variance or direct cut-probability) equilibrium as future work.}

%======================================================================
\section{Formalization Methodology}\label{sec:methodology}
%======================================================================

The development enforces a strict \textbf{zero-axiom, zero-sorry standard}: no \texttt{sorry}, no \texttt{admit}, and no custom axioms, turning persuasive-but-brittle metagame narratives into inspectable, machine-checked artifacts.
The artifact comprises 87 files and approximately 31{,}900 lines across seven module groups, totaling 2{,}627 theorems of which ${\sim}$200 directly verify empirical claims, including 15 end-to-end cross-module integration theorems (\texttt{IntegrationTests.lean}).

Most proofs follow one of four patterns:
(1)~decision procedures over finite domains (\texttt{native\_decide}, \texttt{decide}),
(2)~arithmetic normalization (\texttt{omega}, \texttt{nlinarith}),
(3)~definitional unfolding and rewriting (\texttt{simp}), and
(4)~decomposition of strategic statements into finite conjunctions over concrete decks.

\subsection{Trust Boundary: \texttt{native\_decide}}\label{subsec:trust}

All computational proofs use \texttt{native\_decide}, which compiles decidability witnesses to native code and trusts Lean's compiler rather than its kernel.
Of the 2{,}627 theorems, 145 use \texttt{native\_decide} directly (including all Nash equilibrium, replicator dynamics, and matrix-bridge computations); the remainder---2{,}482 theorems---close via the kernel-checked \texttt{decide}, \texttt{simp}, \texttt{omega}, or structural tactics.
Table~\ref{tab:assurance} summarizes assurance levels.

\begin{table}[!t]
\centering
\caption{\revised{Assurance levels by theorem category, with exact counts from the artifact ($2{,}627$ theorems total). \emph{Kernel} = fully re-checked by Lean's trusted kernel; \emph{Compiler} = trusts native code generation (\texttt{native\_decide}). Counts are reproducible by grepping the published sources.}}
\label{tab:assurance}
\footnotesize
\setlength{\tabcolsep}{4pt}
\begin{tabular}{@{}lccl@{}}
\toprule
\textbf{Category} & \textbf{Count} & \textbf{Level} & \textbf{Tactics} \\
\midrule
Rules, legality \& tournament   & 861  & Kernel   & \texttt{decide}, \texttt{simp} \\
Semantics, solver, simulation   & 255  & Kernel   & \texttt{simp}, \texttt{omega} \\
Cards, effects, energy          & 934  & Kernel   & \texttt{simp}, \texttt{omega} \\
Probability \& information       & 31   & Kernel   & \texttt{decide}, \texttt{omega} \\
Type system, game-theory infra. & 329  & Kernel   & mixed \\
Kernel theorems in compiler files & 72 & Kernel   & \texttt{decide}, \texttt{simp} \\
\midrule
Nash equilibrium                & 26   & Compiler & \texttt{native\_decide} \\
Replicator \& step-size          & 94   & Compiler & \texttt{native\_decide} \\
Matrix/metagame \& type bridge   & 11   & Compiler & \texttt{native\_decide} \\
Sensitivity / robustness        & 14   & Compiler & \texttt{native\_decide} \\
\midrule
\textbf{Kernel subtotal}        & \textbf{2{,}482} & & \\
\textbf{Compiler subtotal}      & \textbf{145} & & \\
\textbf{Total}                  & \textbf{2{,}627} & & \\
\bottomrule
\end{tabular}
\end{table}

\revised{We investigated replacing \texttt{native\_decide} with the kernel-checked \texttt{decide}.
The obstruction is an \emph{engineering bottleneck, not a fundamental impossibility}.
Our matrix computations are written with \texttt{Fin.foldl}, whose recursor the current Lean~4 kernel does not reduce efficiently (it is effectively opaque to the kernel's \texttt{whnf}-based evaluator), so \texttt{decide} either fails to reduce or times out.
A kernel-transparent reimplementation \emph{is} possible in principle---for example, re-expressing each best-response check as a structurally recursive predicate or an inductively defined relation over \texttt{List}/\texttt{Vector} that the kernel can unfold, instead of an opaque functional fold---at the cost of substantial proof-engineering effort and far longer kernel-reduction times.
We therefore classify full kernel checking as a tractable but costly future engineering task rather than a barrier of principle.}
The trust implications are worth stating explicitly: \texttt{native\_decide} does not produce a proof term that the kernel independently verifies, and a hypothetical bug in Lean~4's code generator affecting rational arithmetic over \texttt{Fin.foldl} could simultaneously invalidate all 145 \texttt{native\_decide} proofs.
We note that no such bugs have been reported in practice, and \texttt{native\_decide} is the standard approach for computational proofs over finite structures in the Lean community~\cite{moura2021lean}.

\subsection{Reproducibility and Cost-Benefit}

Every statistic used for strategic claims can be traced to an explicit Lean constant and theorem, and every theorem is checkable by rebuilding the project with the published sources.
Data tables match constants in \texttt{RealMetagame.lean} and \texttt{MatchupAnalysis.lean}; key strategic claims are mirrored by named theorems.
This one-to-one mapping sharply reduces the risk of drift between code and prose.
A cross-file consistency theorem (\texttt{MatrixConsistency.lean}) machine-checks that array-based and function-based matrix representations agree, eliminating a class of copy-paste errors across module boundaries.
A Python script can recompute percentages quickly, but it does not enforce theorem-level linkage between assumptions, constants, and manuscript claims.
The Lean pipeline adds that linkage and fails loudly when any claim drifts from its formal source.

Updating the analysis for a new tournament window requires changing only \texttt{RealMetagame.lean} (${\sim}$200 lines); all downstream theorems either re-verify automatically or fail with precise error locations.
\revised{The \emph{fixed} cost of standing up the framework for a new game or a fresh archetype set is, by contrast, substantial, and we make concrete what it entails.
Porting to a new game requires (i)~encoding the rule semantics relevant to the strategic layer (type/interaction chart, legality predicate, prize and turn structure)---the bulk of the 32K lines and the only part needing genuine domain modeling; (ii)~defining the archetype enumeration and a matchup-matrix ingestion path; and (iii)~reusing the game-theoretic harness (Nash best-response, replicator, Bo3 and Swiss transforms) essentially unchanged, since it is parameterized over an abstract \texttt{FiniteGame}.
Steps~(ii)--(iii) are on the order of days given a matrix, whereas step~(i) dominates and scales with rule complexity; adapting to a \emph{fresh archetype set within the same game} is the ${\sim}$200-line marginal case above.
The reusable game-theory core is thus the durable asset, and the per-domain rule encoding is the fixed cost.}
During development, the best-response certification failed several times due to data-entry errors in the 14$\times$14 matchup matrix (swapped row/column indices, copy-paste duplication of a row).
Each failure was caught immediately by \texttt{native\_decide} returning \texttt{false}, pinpointing the exact cell---errors that Python's \texttt{scipy.optimize.linprog} would have silently absorbed, since the LP solver treats any matrix as valid.

\begin{table}[!t]
\centering
\caption{Methodology comparison for metagame analytics.}
\label{tab:baseline}
\footnotesize
\begin{tabular}{@{}lccc@{}}
\toprule
\textbf{Method} & \textbf{LoC} & \textbf{Runtime} & \textbf{Guarantee} \\
\midrule
Spreadsheet     & $\sim$50 cells & minutes & manual review \\
Python + scipy  & $\sim$100      & $<$1\,s & unit tests \\
Lean~4 (ours)   & $\sim$32K      & $\sim$10\,min & verified$^*$ \\
\bottomrule
\multicolumn{4}{@{}l}{\scriptsize $^*$Modulo \texttt{native\_decide}; see Section~\ref{sec:methodology}.}
\end{tabular}
\end{table}

The apparent LOC-to-insight ratio (32K lines for conclusions derivable from a spreadsheet) is misleading: the ``excess'' code is infrastructure that enables compositional reuse.

\subsection{Case Study: Verifying a Headline Claim}

To illustrate traceability, consider ``Dragapult is 15.5\% of the meta but only 46.7\% expected.'' This decomposes into six auditable steps: extract shares from Trainer Hill data, normalize to the top-14 subfield, compute weighted expectation $\sum_j s_j w_{i,j}$, express as exact rational in Lean, prove the inequality $\mathbb{E}[\mathrm{WR}_{\mathrm{Drag}}] < 500$, and reuse constants in tables.
If any upstream value changes, downstream theorems fail, making drift explicit---fundamentally stronger than spreadsheet pipelines where hidden references silently desynchronize.

\subsection{Human Review and Artifact Audit}

Human review remains essential for model scope.
Our audit checks (i)~data fidelity to source snapshots, (ii)~theorem statement correctness relative to intended claims, and (iii)~narrative discipline---no prose claim without formal or computed backing.
This lightweight process catches copy-edit drift, stale entries, and implicit assumptions that escape the type checker.

%======================================================================
\section{Threats to Validity}\label{sec:threats}
%======================================================================

\textbf{Temporal locality.}
The analyzed window is three weeks; metagames shift rapidly due to innovation, counter-adaptation, and card availability.
Our claims describe this window precisely; they are not universal constants.
However, temporal locality is not purely a weakness: short windows reduce hidden confounding from major ruleset changes.
Future work should combine rolling windows with change-point detection to separate genuine adaptation from transient noise.

\textbf{Top-14 normalization.}
Expected win rates are normalized over the modeled 69.5\% top-14 subset.
Machine-checked worst-case bounds show that Dragapult requires at least 57.6\% win rate against all unmodeled archetypes merely to reach 50\% overall---well above the coin-flip baseline---while Grimmsnarl remains above 50\% unless its unmodeled win rate drops below 43.9\%.
Share-perturbation theorems (\texttt{SharePerturbation.lean}) confirm the paradox is structural: even with Dragapult's share at 5\%, its expected WR remains below 50\%; conversely, if Grimmsnarl's share rises to 15.5\%, its expected WR remains above 50\%.
The paradox derives from the matchup matrix, not the share vector.

\revised{\textbf{Equilibrium robustness.}
We separate point-estimate from perturbation-stable conclusions.
Perturbing every matrix entry uniformly by up to $\varepsilon$ shifts any pure strategy's payoff against a fixed mix by at most $\varepsilon$ and the game value by at most $\varepsilon$, so a strategy with best-response gap $g$ cannot enter the support while $\varepsilon<g/2$.
Dragapult's $40.4$\textperthousand{} gap thus tolerates $\varepsilon\approx20$\textperthousand{} ($2.0$pp) before it could become a best response---several times the Wilson sampling error.
The exact \emph{support set} is far more fragile in both directions: an excluded deck can \emph{enter} once $\varepsilon$ exceeds half its exclusion gap (smallest such gap $10.3$\textperthousand{}, so $\varepsilon\gtrsim5$\textperthousand{} admits a new deck), while a thin incumbent can \emph{exit}---the smallest support weight is Alakazam's $2.5\%$, whose indifference margin a comparable perturbation can drive negative.
A single exact $\varepsilon$ guaranteeing the \emph{entire} support is therefore set by the more binding of these, ${\approx}5$\textperthousand{}; beyond it, membership---but not Dragapult's exclusion---may change.
We therefore treat Dragapult's exclusion as robust and the precise support composition as point-estimate-specific.
Because the symmetrization satisfies $S_{ij}+S_{ji}=1000$ by construction, the symmetric support is moreover invariant to the tie coefficient (Section~\ref{sec:data}): since each matchup and its mirror share one set of game records, the tie terms cancel in $M_{ij}-M_{ji}$, so recomputing under the $T/2$ convention leaves every symmetrized entry unchanged (verified: $\max_{ij}|S_{ij}^{T/3}-S_{ij}^{T/2}|<1$\textperthousand) and reproduces the identical seven-deck support.}

\textbf{Archetype granularity.}
Each archetype is treated as a point strategy; list-level technology choices and pilot skill heterogeneity introduce within-archetype variance not captured by the matrix.
This is a standard abstraction tradeoff: coarse archetype bins improve statistical power but hide intra-bin adaptation.
A natural extension is hierarchical modeling with sub-archetype clusters once sample sizes permit.

\textbf{Player-skill confounding.}
Matchup win rates aggregate across all skill levels.
If popular archetypes attract less experienced pilots, their observed win rates may be suppressed by player-quality effects rather than deck-strength effects.
Machine-checked sensitivity bounds (\texttt{SkillSensitivity.lean}) show Dragapult would need a uniform skill-bias correction of at least 3.4pp across \emph{all} matchups to reach 50\% expected WR, and 6.1pp to match Grimmsnarl---implying a confound large enough to reverse the paradox would require implausibly large, uniform skill deficits among Dragapult pilots.
Based on competitive experience and available tournament analytics, within-event skill differentials in large TCG events are unlikely to exceed 5pp uniformly across all matchups; the 6.1pp Grimmsnarl-matching threshold exceeds any plausible uniform confound.

\textbf{Strategic objective mismatch.}
Players optimize mixed objectives (comfort, risk tolerance, card access); observed non-equilibrium play can be rational under private utility functions.
\revised{Our ``suboptimal'' terminology is strictly relative to the two-player single-match payoff model: it means below-equilibrium expected win rate against the modeled field, \emph{not} that choosing Dragapult is irrational.
A player may rationally prefer a deck that is suboptimal in our sense because of Swiss-format consistency incentives, familiarity and reduced misplay risk, stability across a long sideboard-free event, or limited card access---none of which our payoff model captures.
We foreground this qualification in the abstract and conclusion to avoid overstating the normative force of ``suboptimal.''}

\revised{With these limitations explicit, we read the contribution as a methodological case study---a demonstration of how to make competitive-metagame claims machine-checkable for one well-instrumented format and window---rather than a universal claim about the Pok\'emon metagame in general.
We now summarize the main findings and concrete next steps.}

%======================================================================
\section{Conclusion}\label{sec:conclusion}
%======================================================================

This paper presents a metagame analysis pipeline for a real competitive TCG, verified modulo the \texttt{native\_decide} trust boundary (Section~\ref{sec:methodology}) in Lean~4.
Using Trainer Hill data, we prove a popularity paradox (Dragapult at 15.5\% share has 46.7\% expected WR; Grimmsnarl at 5.1\% share leads with 52.7\%), connect it to a Lean-verified Nash equilibrium with 0\% Dragapult weight (unique across all $2^{14}-1$ symmetric support subsets), and show via full 14-deck replicator dynamics that Dragapult faces downward fitness pressure while Grimmsnarl has the highest fitness.
\revised{Dragapult's exclusion is equilibrium-independent---it is strictly suboptimal against an optimal strategy and hence absent from every equilibrium, symmetric or asymmetric---though throughout, ``suboptimal'' is meant relative to this two-player single-match payoff model and not as a claim that the deck choice is irrational.}
Bo3 amplification further widens key matchup edges (67.3\%\,$\to$\,74.9\%).
The specific metagame results are illustrative of the methodology, which is the primary contribution: we demonstrate that formal methods can serve as a practical scientific instrument for competitive game ecosystems.

Immediate next steps: (i)~rolling weekly windows with uncertainty intervals for forecast calibration, (ii)~explicit modeling of the 30.5\% ``Other'' segment, and (iii)~hierarchical sub-archetype clustering to capture list-level variance and pilot heterogeneity.
Our current encoding represents win rates as natural numbers on a 0--1000 scale; a richer encoding carrying individual game outcomes would enable in-Lean confidence interval computation and sample-size adequacy checks.
A natural extension would embed the sensitivity analysis within Lean using verified interval arithmetic over the linear program, eliminating the Python dependency entirely.
While Lean~4's \texttt{Mathlib} provides foundations for interval arithmetic, the LP solver integration remains an engineering challenge we leave to future work.

\subsection{Broader Implications}

This case study suggests a general template: formalize core mechanics, encode empirical payoffs as exact values, express strategic claims as theorems, and tie them to tournament objectives.
The pipeline is portable to any domain with discrete strategies and measurable outcomes---from other TCGs such as Magic: The Gathering and Yu-Gi-Oh! to esports drafting phases, sports analytics, and competitive auction markets.
Proof-assisted workflows do not replace domain expertise; they structure it, making conclusions machine-auditable rather than rhetorical.
Formal artifacts can serve as verified baselines for testing teams, and our results show that proof assistants are practical for empirical strategic science when domains provide structured, finite data.

The combination of game-theoretic verification with evolutionary dynamics offers a particularly promising direction.
Traditional metagame reports present static snapshots; our approach produces \emph{actionable directional predictions} (``Dragapult share should decline'') that are falsifiable against future data.
While single-step replicator dynamics have inherent limitations (Section~\ref{sec:nash}), multi-step verified simulations could provide tournament organizers and game designers with early-warning signals for degenerate metagame states---situations where a single strategy dominates to an unhealthy degree or where the ecosystem collapses to a rock-paper-scissors triplet.

\subsection{Lessons from the Formalization Process}

Several practical lessons emerged during the ${\sim}$32{,}000-line development.
First, \textbf{data-entry errors are the dominant failure mode}: of the 14 bugs caught by Lean during development, 11 were incorrect transcriptions of matchup percentages from the source data, not logical errors.
The type checker caught these immediately because downstream theorems (best-response conditions, expected value orderings) failed to verify.
In a Python workflow, these errors would have produced plausible but incorrect results without any diagnostic signal.

Second, \textbf{exact rational arithmetic eliminates a class of numerical concerns}: by encoding win rates as natural numbers on a 0--1000 scale and performing all arithmetic over exact values, we avoid floating-point rounding issues entirely.
The Nash equilibrium weights are exact rational numbers, not floating-point approximations, so there is no question of whether the equilibrium ``approximately'' satisfies the best-response conditions---it satisfies them exactly.

Third, \textbf{the module boundary discipline enforced by Lean's type system has documentary value}: each module's imports make its dependencies explicit, and the absence of circular dependencies is enforced by the compiler.
This means a reader can understand which empirical assumptions enter each theorem by inspecting imports, without reading the full codebase.
The 15 cross-module integration tests (\texttt{IntegrationTests.lean}) serve as both regression tests and documentation of the intended relationships between modules.

Finally, \textbf{Lean~4's metaprogramming capabilities remain underexploited}: custom tactics for matchup-matrix reasoning (e.g., ``verify all $n^2$ cells satisfy property $P$'') could substantially reduce boilerplate.
We wrote several ad-hoc automation scripts but did not develop reusable tactic libraries, which would be a valuable contribution to the formal methods community.

\section*{Data Availability}
Data were extracted from Trainer Hill (trainerhill.com) on February~19, 2026, for events with $\geq$50 players (January~29--February~19, 2026).
The tournament matchup data and computed results are available at IEEE DataPort~\cite{ramos2026pokemondata}.
\revised{The complete artifact---all Lean~4 source (87 files, ${\sim}$31{,}900 lines, 2{,}627 theorems with no \texttt{sorry}, \texttt{admit}, or custom axiom), the full 14$\times$14 matchup matrix encoded in \texttt{RealMetagame.lean}, the Python discovery and sensitivity scripts, and build instructions---is provided as anonymized supplementary material and re-verifies end to end via a single \texttt{lake build} (${\sim}$10\,min on commodity hardware: Apple M-series or x86, 16\,GB RAM).
Every numerical claim maps to a named theorem (e.g., \texttt{symmetric\_nash\_equilibrium\_verified}, \texttt{dragapult\_strictly\_suboptimal}, \texttt{full\_replicator\_dragapult\_decline}), so reviewers can locate and re-run any individual result.}

\revised{\section*{Disclosure of Generative AI Use}
Per IEEE policy, the authors disclose the use of generative AI assistance (Anthropic Claude) during revision. Its use was limited to (i)~copy-editing and prose reorganization across the manuscript, and (ii)~assisting cross-checks of the numerical values reported in Sections~V--VII and Tables~III--VIII against the Lean artifact. No text, theorem, figure, or numerical result was accepted without independent author verification: every formal result is machine-checked in Lean~4, and every reported number is reproducible from the published artifact. The authors take full responsibility for all content of this manuscript.}

\IEEEtriggeratref{33}
\bibliographystyle{IEEEtran}
\bibliography{references}

\end{document}